\documentclass[a4paper,twocolumn,11pt,accepted=2024-07-19,superscriptaddress]{quantumarticle}
\pdfoutput=1
\usepackage{amssymb,amsmath,amsthm,bbm,dsfont}
\usepackage{babel}
\usepackage{color}
\usepackage{graphicx}
\usepackage[svgnames]{xcolor}
\usepackage{tikz,pgf}
\usepackage{blkarray} % block matrices
\usepackage[symbol]{footmisc} % for notes in author field
\usetikzlibrary{backgrounds,decorations.pathreplacing,arrows,decorations.pathmorphing,decorations.shapes,plotmarks,decorations.markings}
\usepackage{orcidlink,hyperref}
\PassOptionsToPackage{compress}{natbib}
\usepackage[numbers]{natbib}
\usepackage[utf8]{inputenc}
\usepackage[T1]{fontenc}
 \newcommand{\ket}[1]{\left|#1\right\rangle}
 \newcommand{\bra}[1]{\left\langle#1\right|}
 
\usepackage[normalem]{ulem}

\begin{document}

\title{Non-Markovianity in High-Dimensional Open Quantum Systems using Next-generation Multicore Optical Fibers}
\author{Santiago Rojas-Rojas \orcidlink{0000-0002-2284-0305}}
\email{santirojas@udec.cl}
\affiliation{Departamento de F\'isica, Universidad de Concepci\'on, casilla 160-C, Concepci\'on, Chile}
\affiliation{Millennium Institute for Research in Optics, Universidad de Concepci\'on, casilla 160-C, Concepci\'on, Chile}
\author{Daniel Mart\'inez \orcidlink{0000-0002-8154-7141}}
\affiliation{Departamento de F\'isica, Universidad de Concepci\'on, casilla 160-C, Concepci\'on, Chile}
\affiliation{Millennium Institute for Research in Optics, Universidad de Concepci\'on, casilla 160-C, Concepci\'on, Chile}
\affiliation{University of Vienna, Faculty of Physics, Vienna Center for Quantum Science and Technology (VCQ), 1090 Vienna, Austria}
\affiliation{Christian Doppler Laboratory for Photonic Quantum Computer, Faculty of Physics, University of Vienna, 1090 Vienna, Austria}
\author{Kei Sawada \orcidlink{0009-0003-2396-681X}}
\affiliation{Departamento de F\'isica, Universidad de Concepci\'on, casilla 160-C, Concepci\'on, Chile}
\affiliation{Millennium Institute for Research in Optics, Universidad de Concepci\'on, casilla 160-C, Concepci\'on, Chile}
\author{Luciano Pereira \orcidlink{0000-0003-1183-2382}}
\affiliation{Instituto de F\'isica Fundamental IFF-CSIC, Calle Serrano 113b, Madrid 28006, Espa\~na}
\author{Stephen P. Walborn \orcidlink{0000-0002-3346-8625}}
\author{Esteban S. G\'omez \orcidlink{0000-0003-3227-9432}}
\affiliation{Departamento de F\'isica, Universidad de Concepci\'on, casilla 160-C, Concepci\'on, Chile}
\affiliation{Millennium Institute for Research in Optics, Universidad de Concepci\'on, casilla 160-C, Concepci\'on, Chile}
\author{Nadja K. Bernardes \orcidlink{0000-0001-6307-411X}}
\affiliation{Departamento de F\'isica, Centro de Ciências Exatas e da Natureza, Universidade Federal de Pernambuco, 50670-901 Recife-PE, Brazil}
\author{Gustavo Lima}
\affiliation{Departamento de F\'isica, Universidad de Concepci\'on, casilla 160-C, Concepci\'on, Chile}
\affiliation{Millennium Institute for Research in Optics, Universidad de Concepci\'on, casilla 160-C, Concepci\'on, Chile}
%\pacs{05.60.Gg, 42.82.Et, 42.50.Dv, 42.50.Ex}

\begin{abstract}
\noindent With the advent of quantum technology, the interest in communication tasks assisted by quantum systems has increased both in academia and industry. Nonetheless, the transmission of a quantum state in real-world scenarios is bounded by environmental noise, so that the quantum channel is an open quantum system. In this work, we study a high-dimensional open quantum system in a multicore optical fiber by characterizing the environmental interaction as quantum operations corresponding to probabilistic phase-flips. The experimental platform is currently state-of-the-art for quantum information processing with multicore fibers. At a given evolution stage we observe a non-Markovian behaviour of the system, which is demonstrated through a proof-of-principle implementation of the Quantum Vault protocol. A better understanding of phase-noise in multicore fibers will improve several real-world communication protocols, since they are a prime candidate to be adopted in future telecom networks.
\end{abstract}

\maketitle
\section{Introduction}

Currently, optical fiber-based communication is the fastest method for information transmission \cite{Maher}, mainly due to the multiple alternatives it offers for multiplexing techniques  \cite{Brackett90,sdm}. One new promising method for increasing fiber information capacity is the space-division multiplexing technique based on Multicore Fibers (MCFs). In this case, more information is sent through the fiber by exploiting the extra cores contained in its cladding \cite{enKlaus1}.  MCFs have allowed transmission rates up to 305 Tb/s through a 19-core MCF \cite{Tbs}, setting a new benchmark for ultrahigh transmission capabilities that largely surpass conventional single-mode fibers \cite{Klaus}. Moreover, it has been recently demonstrated that MCFs are compatible with quantum information processing (QI). For instance, high-dimensional quantum cryptographic protocols have been performed using 4-core fibers.  In this case, high-dimensional quantum states (hereafter qudits) are encoded by exploiting the available core modes for single photon propagation in the MCF \cite{Canas0.3,Dynes:16}. This encoding strategy has been further extended into other QI protocols \cite{xavier20,carine20,carine21,taddei2021computational,martinez2022certification}, and to build high-dimensional entangled photons sources \cite{Lee:17,Lee:19,gomez2021multidimensional}. 

In real large-scale networks, optical fibers are exposed to perturbations induced by environmental noise. As a result of these perturbations, information loss could be introduced according to the Markov hypothesis behind noise processes, spoiling the information transmission \cite{Nielsen}. The Markovian nature of noise has significant consequences for quantum communication tasks and can be witnessed by a monotonic decay of channel capacities along the propagation \cite{caps}. At the same time, the interest in quantum dynamics deviating from the Markov hypothesis, namely non-Markovianity (NM), has increased in recent years due to its theoretical relevance and possible application in the protection and processing of quantum information \cite{Wolf,BHP,Rivas_2014,cv1,cv2}. The behavior of non-Markovian processes has been observed in photonic experimental simulations of environmental effects through different schemes \cite{fanchini14,haseli14,coll15,Alvaro15,Alvaro19,Salles,Marques,PhysRevLett.113.240501,Ringbauer,Valencia,entit,limasilva20}. To date, the use of NM as a resource for QI has been mainly focused on entanglement-based protocols \cite{Huelga,spinstar}, while a formal resource theory for NM is still in early development \cite{anand2019quantifying, resource, Bhattacharya_2021}. Thus, a relevant goal to achieve is the experimental observation of NM in a simple scenario without considering entanglement, envisaging its application for QI tasks in the prepare-and-measure scenario. For instance, measuring non-Markovian effects through an adequately defined quantum channel capacity allows linking NM with the efficiency of a specific QI protocol \cite{caps,cohmeas}. Of particular interest is Ref. \cite{qv} that has introduced the Quantum Vault (QV), a protocol for storing and retrieving information encoded in a quantum system subject to non-Markovian evolution. 

In this work, we introduce a new model to characterize the phase-noise of multicore optical fibers, which is arguably the main effect that leads to the degradation of their information capacity. In our model, the environmental interaction is treated as quantum operations corresponding to probabilistic phase-flips acting on path qudits, which are encoded in terms of the core modes available for photon propagation. The types of phase flips considered in the model can readily be changed moving from simplified scenarios to unrestricted ones. As an initial investigation with this model, we focus on the specific conditions under which the phase-noise will lead to a NM map implemented in the path qudits. The dynamics are experimentally realized by controlling the occurrence probability of the desired error operations. The setup adopted is based on a platform recently introduced to control qudit states propagating over MCFs \cite{carine20}. We observe a non-monotonic behavior of three different capacities, allowing a proof-of-principle realization of a QV \cite{qdh,pqc,noisyqdh}. As discussed above, the observation of NM dynamics in a prepare and measure setup configuration is per se interesting, since it may lead to the development of new QI protocols related to data hiding. Nonetheless, we highlight that our model is far more general, and has been developed with the intent of modeling the imperfections to be observed in installed cables of multicore fibers in future high-capacity telecom networks. 

\begin{figure}[!t]
\centering 
\includegraphics[width=0.45\textwidth]{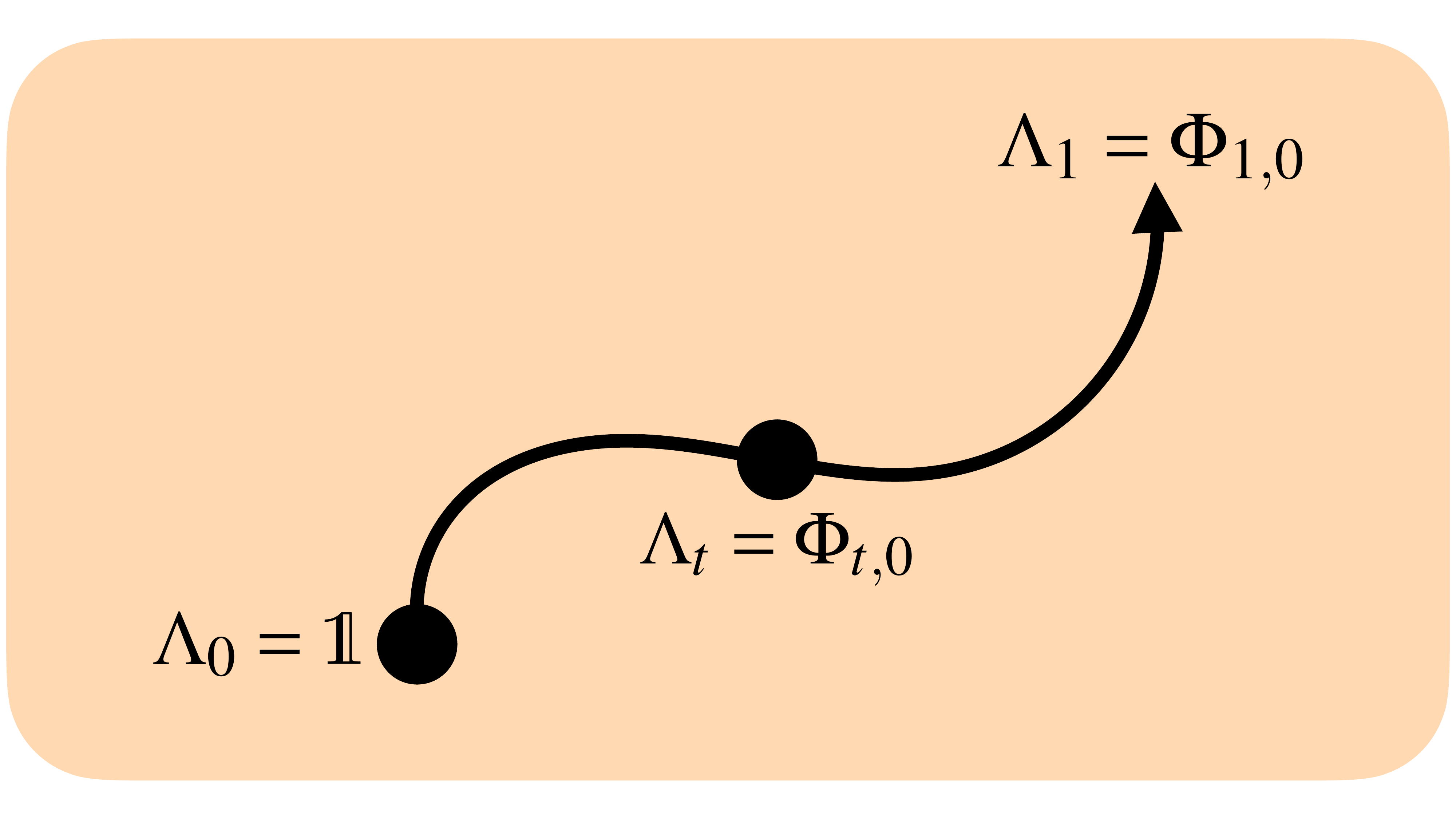}
\caption{A quantum map $\Lambda_t$ can be defined as a curve in the space of quantum channels. We use $t=1-p_0$ as the dynamical variable, where $p_0$ is the no-change probability. Assuming this probability to be null at the end of the evolution, the ending point corresponds to $t=1$.}\label{fg:map}
\end{figure}

\section{Noisy quantum maps in MCFs\label{sec:fun}}
A general quantum process is defined by a family of one parameter dynamical maps $\{\Phi_{t,0}\}$, where $\Phi_{t,0}$ is a completely positive and trace preserving (CPTP) map for any $t>0$ \cite{caps,Breuer1}, taking an input density operator $\rho_0$ into an output density operator $\rho_t$. As depicted in Fig. \ref{fg:map}, a quantum process can be described by a continuous evolution curve $\Lambda_t$ in the space of CPTP maps, such that $\Lambda_0=\mathds{1}$. An evolution is called CP-divisible if $\Phi_{t,0}=\Phi_{t,s}\Phi_{s,0}$, where $\Phi_{t,s}$ is completely positive (CP) for all stages $s$ such that $0\leq s \leq t$. All processes given by CP-divisible maps are called Markovian evolutions \cite{Rivas_2014}. Otherwise, they are called non-Markovian. 

In the following, we consider a noisy quantum process where a quantum system in an unknown state may be affected by a set of unitary transformations $U_k$ operating with their respective probability $p_k$. The formalism of quantum processes allows us to define such a dynamical map following an intuitive posit: when the system is just starting to evolve from its initial state $\rho$, it does not undergo any change yet, so the probability $p_0$ of it remaining unchanged (or being transformed only by $U_0={\mathds{1}}$) must be equal to one, with all the remaining probabilities $p_{k\neq 0}$  being null; as it evolves, the no-change probability $p_0$ decreases while the remaining probabilities increase. (A similar approach has been used for the experimental simulation of environmental effects on single qubits \cite{Salles}.) This way, we associate the dynamical variable $t$ with a decay of $p_0$. Since all the relevant behavior for our study is observed for a finite value of $p_0$, knowing the particular dependence of $t$ on it is not crucial. Then, we can take $t=1-p_0$ as the dynamical parameter to define the noisy process, which is now enclosed by the curve in $0\leq t\leq 1$ or $1\geq p_0\geq 0$.

\begin{figure}[!t]
\centering
 \includegraphics[width=0.45\textwidth]{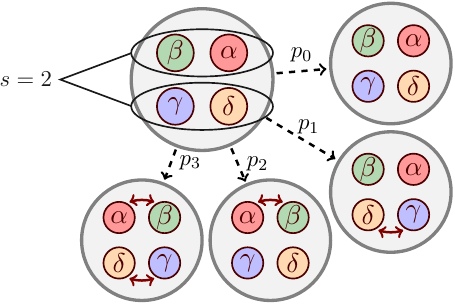}
 \caption{The four possible operations when $N=4$ and $s=2$, with their respective probabilities $p_i$. Greek symbols represent the phases of the logical state at each core. \label{fg:interfaces}}
\end{figure}

The explicit definition of our map makes use of the \emph{operator-sum} representation for a quantum channel $\Phi(\rho)=\rho'$ in the framework of quantum operations, where all the environmental effects are captured by a complete set of \emph{Kraus operators} $E_k$ acting on the system's Hilbert space as $\rho'=\Phi(\rho)=\sum_k E_k\rho E_k^\dagger$. The noisy process $\Lambda_t$ under our consideration is composed of a family of channels whose Kraus operators $E_i=\sqrt{p_i}U_i$ describe the occurrence of a unitary error operation $U_i$ [see the examples below in \eqref{eq:n4s2}] affecting the system with probability $p_i$, all the $p_i$ being defined in function of the no-change probability $p_0$. Let $\ket{\psi}$ be the state of  a single photon propagating through a $N$-core optical fiber: 
\begin{equation}\label{eq:state}
 \ket{\psi}=\frac{1}{\sqrt{N}}\sum\limits_{\ell=1}^Ne^{i\phi_\ell}\ket{\ell}\,,
\end{equation}
where $\ket{\ell}$ denotes a one-photon state in the core $\ell$. We will consider a set of error operations $U_i$ that permute the phase values  $\phi_\ell$ of the state in Eq. \eqref{eq:state}.  Each permutation on its own is a noiseless operation, so the presence of noise is due to the probabilistic combination of them and somewhat extends the concept of bit-flip or phase-flip channels found in open qubit systems to higher dimensions. We can address different scenarios by allowing permutations only between cores on certain subsets (this has some relation to the prescription given in \cite{sudoku,rsudoku} to implement non-Markovian phase patterns). Let $s$ be the number of cores in these subsets: $s=1$ prevents reordering, so the only possible unitary transformation is the identity, leading to trivial Markovian dynamics. The $s=2$ case (Fig. \ref{fg:interfaces}) involves just  four possible operations, namely,
%\begin{equation}\label{eq:n4s2}
%\begin{split}
% U_0&=\mathds{1}_4\,\quad \quad \quad \quad \quad \quad \quad
% U_1=
% \begin{blockarray}{c|c}
%  \begin{block}{(c|c)}
%  \, \mathds{1}_2 \quad & 0 \\ \BAhline \, 0 \quad & \begin{array}{cc} 0 & 1 \\ 1 & %0 \end{array} \\
%  \end{block}
% \end{blockarray}\;\;,
% \\ 
% U_2& =
% \begin{blockarray}{c|c}
%  \begin{block}{(c|c)}
%  \begin{array}{cc} 0 & 1 \\ 1 & 0 \end{array} & \,  0 \quad
%  \\ \BAhline
%  0 & \, \mathds{1}_2 \quad   \\
%  \end{block}
% \end{blockarray} \;\;,
% U_3 =\;
% \begin{blockarray}{c|c}
%  \begin{block}{(c|c)}
%  \begin{array}{cc} 0 & 1 \\ 1 & 0 \end{array} & \, 0 \quad
%  \\ \BAhline 
%  \, 0 \quad & \begin{array}{cc} 0 & 1 \\ 1 & 0 \end{array} \\
%  \end{block}
% \end{blockarray}\,,
%\end{split}
%\end{equation}
\begin{equation}\label{eq:n4s2}
\begin{split}
 U_0=\begin{pmatrix} 1 & 0 & 0 & 0\\
 0 & 1 & 0 & 0\\
 0 & 0 & 1 & 0\\
 0 & 0 & 0 & 1\end{pmatrix},
&\ 
 U_1=\begin{pmatrix} 1 & 0 & 0 & 0\\
 0 & 1 & 0 & 0\\
 0 & 0 & 0 & 1\\
 0 & 0 & 1 & 0\end{pmatrix},
 \\[1ex]
 U_2=\begin{pmatrix} 0 & 1 & 0 & 0\\
 1 & 0 & 0 & 0\\
 0 & 0 & 1 & 0\\
 0 & 0 & 0 & 1\end{pmatrix},
&\ 
 U_3=\begin{pmatrix} 0 & 1 & 0 & 0\\
 1 & 0 & 0 & 0\\
 0 & 0 & 0 & 1\\
 0 & 0 & 1 & 0\end{pmatrix},
\end{split}
\end{equation}
while $s=3$ and $s=4$ enable 6 and 24 unitary transformations, respectively, corresponding to the number of available permutations. It is worth remarking that instead of the most common method used in an experimental simulation of non-Markovian processes --- that is, to consider and manipulate the state of the environment explicitly --- our approach reflects the environmental influence by the variation on the specific probability of an error (permutation) occurring, similar to the scheme used in Ref. \cite{Marques} to implement a dephasing channel or the simulation of classical non-Markovianity in \cite{entit}. In this way, we intend to make our experimental simulation consistent with the standard premise that only the state of the system is accessible to the experimenter, so it can be used as a building block to describe a general noise process for qudits, just like in qubit systems the bit-flip and phase-flip errors are components of the general depolarizing channel. The inaccessibility of the environment is also relevant for using our system as a secure channel.

\section{Non-Markovian effects in quantum information capacities}\label{sec:qi}
\begin{figure}[!tb]
\centering
 \includegraphics[width=0.45\textwidth]{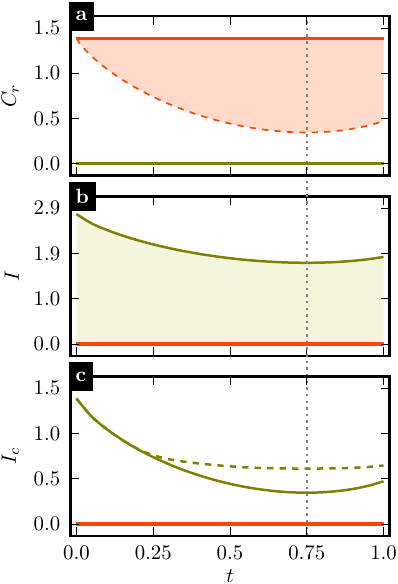}
 \caption{Evolution of coherence and quantum mutual information under maps $\Lambda_{p_0}$ for $s=2$ and when permutations are equally probable (See Fig. \ref{fg:interfaces}). (a) Relative entropy of coherence as a function of $t$ from maximally coherent initial states. If the relative phases between the terms of the state are null, its REC is constant, as depicted by the red line. The other cases are contained in the orange region. (b) Quantum mutual information at each stage of the map $\Lambda_t$. The beige region contains the results obtained from 10$^6$ normally-distributed mixed states. Green solid curves correspond to a chaotic input state ${\mathds{1}}/4$. (c) Evolution of coherent information. In this case, we show the maximum at each point obtained from a normal distribution of 10$^6$ initial mixed states (dashed line). Green solid curves correspond to a chaotic input state.  The vertical dashed lines indicate where the minimum is reached.}\label{fg:caps}
\end{figure}

In order to assess the effect --- and usefulness --- of NM in quantum information and quantum communication protocols, different  NM measures have been proposed, which are based on the monotonic decay of certain quantifiers under the action of CPTP maps. An increase or revival of these quantities along evolution indicates a backflow of information from the environment to the system \cite{BHP,Rivas_2014}. That is, when these quantities show an increase or resurgence during evolution, it indicates a transfer of information from the environment back to the system, serving as evidence/witness of NM's evolution. 

The first information resource we can consider is the \emph{quantum coherence}. If a system evolves under a CPTP map, such as $\Lambda_t$, its coherence cannot increase. Conversely, coherence is expected to decay monotonically under a Markovian noisy process, so a suitable NM witness is given by an increase in the \emph{relative entropy of coherence} (REC) \cite{qirec, cohmeas}, which has a closed form for a $N$-dimensional quantum state $\rho$ given by:
\begin{equation}
 C_r=S(\rho_{\rm diag})-S(\rho)\,.
\end{equation}
Here, $\rho_{\rm diag}$ is the diagonal part of the density matrix $\rho$, and $S$ denotes the von-Neumann entropy \cite{vonNeumann}. This relation with the entropic characteristics of the system poses the maximum REC as a valid measure of the channel capacity, providing an interpretation of coherence in terms of the average amount of information conveyed by a state. We evaluate its behavior under our map $\Lambda_t$, taking as initial states the set of maximally coherent pure states with $C_r=\log 4$.  The orange region depicts the respective evolution of $C_r$ in Fig. \ref{fg:caps} (a) for $s=2$ when the error operations are equally probable, i.e., $p_{i\neq 0}=(1-p_0)/3$. We observe how $C_r$ decays to a minimum at a certain $t$ corresponding to a finite no-change probability $p_0$, after which it increases. The minimum is reached when all the permutations are equally probable, i.e., when $p_i=p_\text{min}$ for all $i$.
The revival of REC in the regime $1-p_\text{min}<t<1$ (or equivalently $p_0<p_\text{min}$) breaks the monotonic decrease imposed by the Markov approximation.

Diagonal density matrices correspond to incoherent states giving a null value of distance-based measures of coherence such as REC [green line in Fig. \ref{fg:caps} (a)]. However, these states have maximal information content that can be transmitted through a quantum channel, which can be measured by proper capacities capturing the \emph{non-unital part} of the dynamics \cite{Rivas_2014}.
In analogy to the classical Shannon capacity, different quantities are used to estimate a bound for the average amount of transmitted information, depending on the particular protocol. First, the \emph{quantum mutual information},
\begin{equation}\label{eq:qI}
 I(\rho,\Phi)=S(\rho)+S(\Phi[\rho])-S(\rho,\Phi)\,,
\end{equation}
limits the amount of classical information that can be transmitted through the channel $\Phi$. In this definition, the last term 
corresponds to the \emph{entropy exchange} $S(\rho,\Phi)$, which quantifies the entropy change with the environment \cite{caps,Holevo}. It is also used to define the \emph{coherent information} \cite{Schumacher}
\begin{equation}\label{eq:Ic}
 I_c(\rho,\Phi)=S(\Phi[\rho])-S(\rho,\Phi)\,.
\end{equation}
This quantity is important for several reasons, which are discussed in more detail in the supplementary material. Its fundamental definition considers the entanglement between the system and an \emph{ancilla}, this being the reason why it is regarded as a proper bound of the nonclassical information conveyed by the quantum carrier \cite{Lloyd}.

Quantum mutual information and coherent information cannot be increased by any post-processing of the channel output. This property, the \emph{data processing inequality} \cite{Marinescu}, is related to the monotonic decay of both capacities under a Markov noisy process. A break of this behavior is a signature of NM evolution \cite{caps}.  In Fig. \ref{fg:caps}, we present the evolution of the quantum mutual information (b) and the coherent information (c) under $\Lambda_t$ for different initial states. Dark green curves correspond to the evolution of the capacities for a mixed input state $\rho=\mathds{1}/d$. Noticeably, both capacities $I$ and $I_c$ attest to the non-Markovian nature of the map: for mixed states, a minimum is again reached at $p_0=p_\text{min}$. According to its definition, the decrease of $I$ (of $I_c$) up to this stage of the evolution indicates a loss of classical (quantum) correlations between the input and the output, with a consequent variation of entropy on the environment, or equivalently, a leak of information from the system to it. In the region $1-p_\text{min}<t<1$, both capacities rise again, indicating an information back-flow from the environment to the system. Note that since all the maximally coherent states of Fig. \ref{fg:caps} (a) are pure, their von Neumann entropy is null, so the related quantum mutual information is also null [red lines in Figs. \ref{fg:caps} (b,c)]. 

\section{Non-Markovian dynamics in MCF}\label{sec:qv}
\subsection{The quantum vault protocol}\label{sec:qvdes}
As far as they account for the maximal average amount of information that can be transmitted, the channel capacities introduced in the previous section identify resources that can be used to store and retrieve information by state preparation and measurement of a quantum state. This is the concept of the quantum vault, originally introduced in Ref. \cite{qv}. Consider that Alice wants to store some information by encoding it in a qudit. The system evolves by the map $\Lambda_t$, with the resource at each stage being quantified by some channel capacity $K(t)$. After the process is finished, i.e., for $t_f=1$ ($p_0=0$), Alice tries to recover her information. The more information that is contained in the final state, the more successful the retrieval is. Suppose that at some stage of the evolution in the interval  $0<t<1$, an eavesdropper Eve attempts to measure the state of the system and thus steal Alice's information. If Eve cannot retrieve as much information as Alice, then the system constitutes a quantum vault \cite{qv}. The noisy process $\Lambda_t$ has a key feature that enables the implementation of the quantum vault, namely, the non-monotonic behavior. There is a finite interval where $K(t)<K(t_f)$, and less information can be retrieved. The revival time $\Delta t$, where the capacities increase at the end of the evolution, can then be used to quantify the suitability of a given system to serve as a QV. 

In a general $N$-core fiber, if $m$ is the integer quotient of $N$ divided by the size of the subsets $s$, there are $(s!)^m\,(N-ms)!$ possible permutations. Since we consider $t_f=1$, the uniform probability $p_\text{min}$ is equal to $\Delta t$ and takes the value
\begin{equation}\label{eq:pmin}
 \Delta t = p_\text{min}=\frac{1}{(s!)^m\,(N-ms)!}\,.
\end{equation} If we consider the error operations to occur with the same probability, i.e. $p_{i\neq 0}=(1-p_0)/3$ in a 4-core fiber with $s=2$, we get $\Delta t=0.25$ (See Fig.~\ref{fg:caps}). However, from Eq. \eqref{eq:pmin}, it is clear that the value of the revival time $\Delta t$ is inversely proportional to the number of available permutations affecting the evolution. Thus, maximal information recovery in the QV protocol can already be obtained while considering a simplified scenario in the $s=2$ case, where phase permutations can only occur simultaneously on both subsets of cores, i.e., $p_1=p_2=0$ and $p_3=1-p_0$. In this case, the revival time is $\Delta t=0.5$. 

\subsection{Experimental Setup}\label{sec:exp}
\begin{figure*}[!t]
\includegraphics[width=\textwidth]{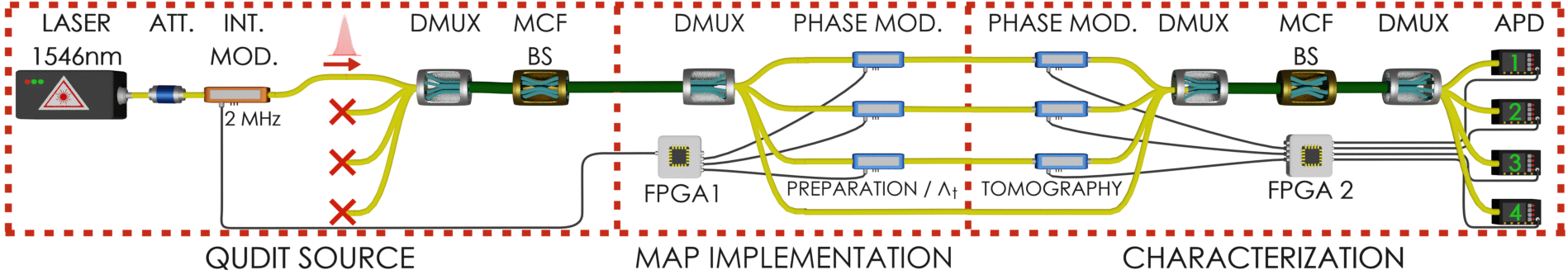}
\caption{\label{fg:setup1col}Four-arm Mach Zehnder interferometer based on MCF to simulate non-markovian dynamics. A CW-1546 nm laser, neutral attenuators, and an intensity modulator generate a weak coherent state with a mean photon number of $\mu=0.15$ at a repetition rate of  2 MHz. The single photon state is transmitted through an MCF DMUX and an MCF BS, generating a 4-dimensional state encoded on the 4 possible core modes inside the MCF. An MCF DMUX is used to spatially separate the paths such that one can address the relative phases via PMs. During a 100 ms time interval, an FPGA-based active control system adjusts the voltages on the PMs to generate the state $\ket{e_i}$. Immediately thereafter, in the next 100 ms time interval,  FPGA1 uses the PMs to randomly implement i) the identity or ii) the phase-flips associated to the transition $3$ in Fig. \ref{fg:interfaces}, with probability $p_3=1-p_0$. In the characterization stage, a second FPGA performs measurements over 5 MUBs to reconstruct the resulting density matrix. This is done by closing the interferometer with a MCF BS and collecting the single photon counts with four APDs.}
\end{figure*}

In order to demonstrate a new platform to study the dynamics of high-dimensional open quantum systems, we exploit the use of a MCF-based 4-arm Mach-Zehnder interferometer to prepare four-dimensional ($d=4$) qudit states, and also to statistically apply a set of unitary operations $U_i$ to them. By doing this, we simulate the state transition maps. The experiment consists of three main parts: state preparation, map implementation, and quantum state tomography characterization, and is presented in Fig. \ref{fg:setup1col}.

\textit{Qudit Source}. A CW-1546 nm laser attenuated by neutral filters and a fiber intensity modulator controlled by a Field Programmable Gate Array (FPGA1) prepares light pulses with a repetition rate of 2 MHz, which are propagated by telecom SMF-28 fibers. The combination of these three devices allows the generation of weak coherent pulses with an average photon number of $\mu=0.15$, which provides a good approximation to a single photon source, since $95.7\%$ of the non-vacuum pulses are single photons. Next, a demultiplexer (DMUX) device is used to connect the initial SMF into a single core of a four-core MCF, while the other three cores remain unconnected. A backbone device in the setup is the MCF Beam Splitter (MCF-BS), which enables the superposition state on the propagation paths of the MCF\cite{carine20,taddei2021computational}. An MCF-BS is fabricated by heating and stretching a section of the MCF, creating a tapered subsection of the fiber. This geometric deformation of the MCF structure changes the distance between the cores, allowing crosstalk between them. The MCF-BS was recently characterized by Cariñe et al. \cite{carine20}, who proved that the output distribution corresponds to a $d=4$ Haddamard matrix $H_4$, allowing the generation of superposition states. Thus, the qudit state after MCF-BS is $\ket{\psi_0} = ( 1/2 ) \sum_{\ell=1}^4 e^{i\phi_\ell} \ket{\ell}$, where $\phi_k$ are individual phase drifts that will be addressed in the next stage.

\textit{Map Implementation.} To extinguish parasitic phase drifts in state preparation and to apply the probabilistic dynamical map, we include a phase stabilization system in the setup. It is composed of three LiNbO$_3$ phase modulators (PMs) in a feedback-loop connected with the avalanche photo detectors (APDs). The control system works as follows: Initially, there is a 100 ms stabilization interval, where FPGA1 implements a stabilization procedure \cite{carine20} to compensate for phase drifts between cores and generate the desired initial state. This is done by optimizing the observed probability distribution in the APDs, such that it is very close to the distribution expected for the state that one wants to generate. Typical fidelities observed in this stage are higher than 99$\%$, due to almost perfect mode matching for the different core modes in the final beam-splitter.

Once the target output probability is reached, the dynamic map $\Lambda_t$ is implemented for the next 100 ms time interval. This is done by randomly applying the unitaries $U_0$ and $U_3$ (see Fig. \ref{fg:interfaces}), for each of the 200.000 pulses generated, with probabilities $p_0$ and  $1-p_0$, respectively.  In this way, the observed statistics, integrated over the 100 ms time interval, are associated with a mixed state that emerges after the evolution through a map $\Lambda_t$, with $t=p_3=1-p_0$.

\textit{Characterization.} A quantum state tomography with five mutually unbiased bases (MUBs) in $d=4$ is used to estimate the final qudit density matrix \cite{MUB1,MUB2,MUB3,MUB4,MUB5,MUB6,MUB7}, allowing the calculation of the required channel capacity. The measurement projection is implemented using a second FPGA2 that controls a second set of phase-modulators and with the final $4 \times 4$ MCF beam-splitter. The states associated with the four outcomes of this last beam-splitter are given by 
\begin{eqnarray}
|\psi_{0}\rangle= \frac{1}{2}(e^{i\phi_{0}^{B}} |0 \rangle+e^{i\phi_{1}^{B}} |1 \rangle+e^{i\phi_{2}^{B}} |2 \rangle+e^{i\phi_{3}^{B}} |3 \rangle) , \nonumber \\
|\psi_{1}\rangle= \frac{1}{2}(e^{i\phi_{0}^{B}} |0 \rangle+  e^{i\phi_{1}^{B}} |1 \rangle- e^{i\phi_{2}^{B}} |2 \rangle- e^{i\phi_{3}^{B}} |3 \rangle), \nonumber \\
|\psi_{2}\rangle= \frac{1}{2}(e^{i\phi_{0}^{B}} |0 \rangle- e^{i\phi_{1}^{B}} |1 \rangle+e^{i\phi_{2}^{B}} |2 \rangle-e^{i\phi_{3}^{B}} |3 \rangle), \nonumber \\
|\psi_{3}\rangle= \frac{1}{2}(e^{i\phi_{0}^{B}} |0 \rangle-  e^{i\phi_{1}^{B}} |1 \rangle- e^{i\phi_{2}^{B}} |2 \rangle+ e^{i\phi_{3}^{B}} |3 \rangle),
\end{eqnarray} where $\phi_k^{B}$ is the phase applied by the second modulator in the core mode $k$. The projection is concluded connecting commercial InGaAs single-photon detection modules to each output mode, working in gated mode and configured with 10\% overall detection efficiency. Over the 100 ms measurement interval, FPGA2 imprints the phases that correspond to the 16 different states of the four MUBS and records the corresponding statistics. The fifth MUB (logical basis) was measured before each round of the experiment. The results were used to compute the experimental density matrix $\rho_{exp}$ using a maximum likelihood estimation routine \cite{Shang2017}, and then the different channel capacities.

\subsection{Proof-of-principle realization of the QV protocol}\label{sec:exp}

\begin{figure*}[!t]
\centering
\includegraphics[width=0.8\textwidth]{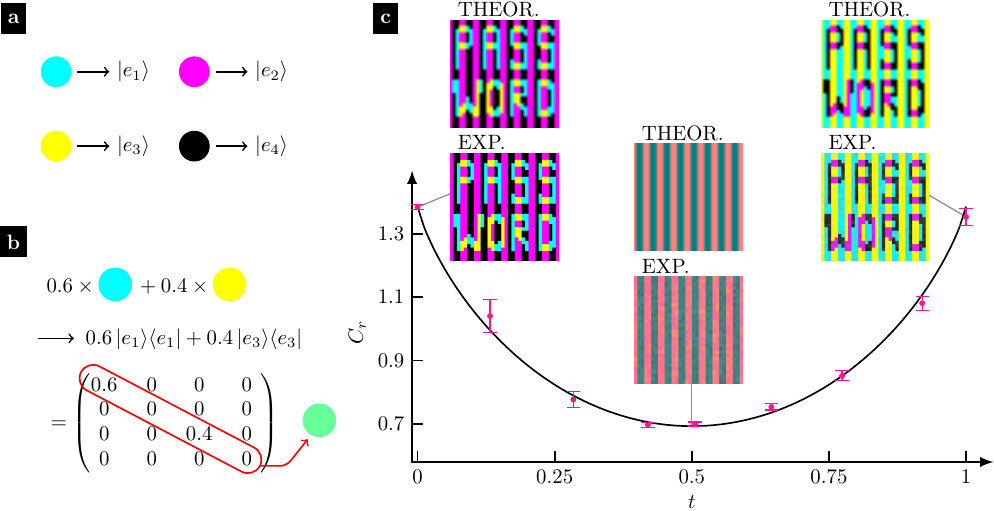}
\caption{Proof-of-principle implementation of the quantum vault protocol to store an image using the noisy map $\Lambda_t$ with $p_1=p_2=0$. (a) Each qudit in the 4-dimensional canonical basis is associated with a specific color of the cyan-magenta-yellow-black (CMYK) model. (b) Mapping the CMYK color model into the basis allows a direct equivalence between color mixtures and mixed states: the diagonal of the density matrix in this basis is the CMYK code of the respective color. (c) At the input stage, Alice stores a message as an image whose pixels have one of the base colors. As the system evolves, the distinguishability between different state pairs in the basis decreases, which is reflected by a decrease in the REC. At $t=1-p_{\rm min}$, the capacity is minimal, rendering the message of the image unreadable. Non-Markovian dynamics in the final stage of the evolution leads to an increase in distinguishability, allowing Alice to retrieve her information even if the final state of each pixel is different from its initial state (colors are swapped in the retrieved image). In the insets, the top and bottom images correspond to the predicted and measured images. (Please see Fig.~\ref{fg:qv_sp} where we present different cases in the supplementary material)}\label{fg:qv}
\end{figure*}

The fact that the randomly applied unitaries are actually performed by the user allows the NM to be exploited in a controlled way. Let us illustrate this by outlining how the QV procedure applied to our MCF setup can be used to hide information efficiently. Suppose the photons are sent through a long MCF fiber that serves as a quantum memory. Let us apply phases using the relevant probability distribution corresponding to $p_{min}$. Then, the information that an eavesdropper can recover is minimal.  Using a classical register, the user can make a record of the particular unitary $U_i$ that was applied to each light pulse sent through the fiber.   When the photons exit the fiber, if one has access to the classical register, one can look up which unitary was applied to each photon and apply a second unitary chosen so that the complete evolution corresponds to a non-Markovian channel with $t=1$, where information recovery is maximum. An eavesdropper who does not have access to the classical register does not know which unitaries have been applied to which photons and thus cannot generate a non-Markovian channel.  In this way,  the information hidden in the vault can only be recovered by users with access to the classical memory register. 

Consider an arbitrary orthonormal base defined by the states:
\begin{equation}
 \begin{split}
  &\ket{e_1}=\frac{1}{2}\begin{pmatrix}  1 \\
 i \\
 i \\
 -1 \end{pmatrix}\,,\quad\ket{e_2}=\frac{1}{2}\begin{pmatrix}  1 \\
 i \\
 -i \\
 1 \end{pmatrix}\,,\\
 &\ket{e_3}=\frac{1}{2}\begin{pmatrix}  1 \\
 -i \\
 -i \\
 -1 \end{pmatrix}\,,\quad
 \ket{e_4}=\frac{1}{2}\begin{pmatrix}  1 \\
 -i \\
 i \\
 1 \end{pmatrix}\, ,
 \end{split}
 \label{eq:mub4}
\end{equation} which are all equally weighted superpostions of  the states $\ket{\ell}$ ($\ell=0,1,2,3$) corresponding to a single photon propagating in the $\ell$-th core. To simulate different channel dynamics in our setup, the evolution of the input state $\ket{e_2}$ was studied. For each measurement interval of 100 ms, we randomly applied $U_0$ and $U_3$ considering different values for the probabilities $p_0$ and $p_3$. The red dots in Fig. \ref{fg:qv} correspond to the evolution of the relative entropy of coherence $C_r$ at specific stages in the simplified case ($p_3=1-p_0$ and $p_1=p_2=0$). This scenario allows $C_r$ to rise back to its initial value at the end of the evolution, so the information backflow is complete. The simulation consists of summing up, for a given stage, the statistics obtained in all previous ones. Thus, simulating the photon state evolution through different channel dynamics, sequentially. The experimental results follow the theoretical prediction very closely, serving as a demonstration that different channel dynamics (Markovian and Non-markovian) can be simulated in a very controlled way in our system. To compute the error bars for the experimental data set, we performed a Monte Carlo routine by varying a Poisson distribution centered on the experimental counts and computed 1000 simulated density matrices. Error bars correspond to one standard deviation of the ensemble.

Now, let us show how a classical message is affected by such different types of dynamics. Consider an image composed of 1024 pixels whose color is defined in the four-color model CMYK (cyan, magenta, yellow, black). This pattern of colors can be encoded directly in the quantum state of the photons sent through our setup. As depicted in Fig. \ref{fg:qv}b, associating the basis states above with the CMYK code, provides a straightforward visual representation of mixed states as color mixtures. The user stores a four-color image whose pixels correspond exactly to one of the four CMYK colors, as shown in the top left inset of Fig. \ref{fg:qv}c. This is done by preparing a set of 1024 qudits, whose states are defined according to the CYMK color code. As the evolution proceeds, each of these pixel-states is independently affected by the noisy map $\Lambda_t$, so they turn into a mixed state $\rho_t$. In Fig. \ref{fg:qv}c, we plot the respective images retrieved. The output image was recovered by measuring in the encoding basis \eqref{eq:mub4}. The full backflow of the information encoded is reflected by the last image, where all pixels are in a pure color of the CMYK code, just like the initial image. The minimum of the capacity $C_r$ is reached with $p_3=p_0=1/2$ ($t=0.5$), when all the pixel-states at this point are in one of two mixtures, rendering the message in the image unreadable. The average fidelity of the reconstructed qudits is $98.5\%$, so the difference between the experimental and theoretical images is almost negligible.  

\section{Conclusions \label{sec:concs}} 
Applying quantum technologies in practical scenarios requires the study of decoherence, which is intrinsically related to environmental noise, that can outweigh the advantages that quantum systems could provide. In this context, studying open quantum systems and non-Markovianity is a relevant topic, as it could provide bright insights for fault-tolerant quantum communication tasks, real-world system evolution, or even practical quantum memories \cite{Mandayam2023,buscemi_information_2024}.

We have shown that multicore optical fibers provide a suitable platform to implement NM maps in higher dimensions through the probabilistic application of unitary operations in each fiber core optical mode.  We used such technique in a proof-of-principle experiment to show that under a restricted scenario for phase flips, if the multicore fiber is used as a quantum memory, NM maps may lead the system evolution to a regime where information backflow from the environment to the system is observed. Thus, enabling the use of the system as a quantum vault where information is harder to retrieve along the evolution than at the end of it.

The set of errors used in our study serves as a reliable proving ground that can be extended (with $s>2$) to include more noisy operations affecting open systems in actual communication networks implemented over installed multicore optical fibers. A further study of transition map models in multicore fibers could be useful for both classical and quantum communication tasks since a deep understanding of the noise contribution could lead to the implementation of noise passive filters, thus reducing the complexity and cost of communication infrastructure networks. 

\section*{Acknowledgments}
\par
This work was supported by Fondo Nacional de Desarrollo Científico y Tecnológico (ANID) (Grants No. 3200779, 1200266, 1231940, 1200859, 1240746) and ANID – Millennium Science Initiative Program – ICN17\_012. K.S. acknowledges financial support from project UCO 1866. LP was supported by ANID-PFCHA/DOCTORADO-BECAS-CHILE/2019-772200275, the CSIC Interdisciplinary Thematic Platform (PTI+) on Quantum Technologies (PTI-QTEP+), and the Proyecto Sinérgico CAM 2020 Y2020/TCS-6545 (NanoQuCo-CM). N.K.B. acknowledges financial support from CAPES, CNPq Brazil (Universal Grant No. 406499/2021-7), and FAPESP (Grant 2021/06035-0). N.K.B. is part of the Brazilian National Institute for Quantum Information (INCT Grant 465469/2014-0).

\bibliographystyle{quantum}
\bibliography{multicore}

\appendix
\section*{Supplementary material}

\subsection*{Coherent information and channel privacy}

We will now discuss in more detail the original definition of coherent information and its relation to the privacy of a quantum channel. Consider that in addition to the original system $A$, we have a replica of it (either physical or just a mathematical device) that serves as a reference or ancilla system $R$.  Consider a quantum channel $\Phi$ which acts only in the system $A$ and describes its interaction with the environment $E$. This channel transforms the initial state $\rho_A$ of $A$ into a final state $\rho'_A$. Generally, $\rho$ is a mixed state but can be purified in the bipartite Hilbert space ${\mathcal H}_A\otimes{\mathcal H}_R$. The purification of $\rho$ is given by the entangled state $\ket{\Psi_{AR}}$. We can express the evolution of the extended system $AR$ by means of the superoperator $\Phi\otimes {\mathds{1}}_R$ to obtain
\begin{equation}
 \rho_{AR}'=\left\lbrace \Phi\otimes {\mathds{1}}_R \right\rbrace (\ket{\Psi_{AR}}\bra{\Psi_{AR}})\,.
\end{equation}

This translates into the following scheme:

\begin{center}
  \begin{tikzpicture}
   \draw[thick] (0,0) node(rhoa)[circle,draw,fill=white]{$\rho_A$} --++(3,0) node(rhoap)[circle,draw,fill=white,inner sep=2.5pt]{$\rho_A'$};
   \draw[thick,latex-latex] (1.5,0) --(1.5,-1) node[below,draw]{$E$} node[midway,right]{$\Phi$};
   \draw[thick,dotted](rhoa) --+(0,1) node[draw,solid,fill=white]{$R$};
  \end{tikzpicture}
 \end{center}

Now, consider the entropy $S(\rho_{AR}')$ of the bipartite system composed of the output and the ancilla. Since this subsystem is not affected by the evolution, $S(\rho_{AR}')$ can still be considered as an intrinsic property of the system $A$, depending only on its initial state $\rho_A$ and the channel $\Phi$ \cite{Schumacher}. Still, if the environment is initially in a pure state, we can identify $S(\rho_{AR}')$ with its entropy after evolution. The previous statements allow us to define the entropy exchange,
\begin{equation}
 S(\rho_A,\phi)=S(\rho_{AR}')=S(\rho_E')\,,
\end{equation}
as the amount of information exchanged between $A$ and the environment. The last identity comes from the fact that the tripartite state in the extended system $ARE$ remains pure along the evolution. Upon the entropy exchange, we define the two intrinsic quantities introduced in Sec. \ref{sec:qi}, namely, the quantum mutual information and the coherent quantum information. This last quantity, given by
\begin{equation}
 I_c(\rho_A,\Phi)=S(\Phi[\rho_A])-S(\rho_A,\phi)\,,
\end{equation}
measures the degree of quantum correlation retained by $R$ and $A$ during the evolution, so it is considered as a proper measure of the nonclassicality of the final state $\rho_{AR}'$. The previous definitions provide a formal connection between our map and the task  of \emph{quantum data hiding} \cite{qdh,pqc}. In this context, removing correlations between $A$ and $R$ allows to hide quantum information by transferring them to the environment (cf. \cite{buscemi_information_2024}). This is quantified by the entropy exchange and the associated reduction of the coherent information. 

Consider the evolution of the coherent information $I_c$ under the map $\Lambda_t$, shown in Fig. \ref{fg:caps} (e,f). This is obtained by taking the chaotic initial state $\rho_A={\mathds{1}}/4$, whose purification is given by a maximally entangled state. As $t$ increases, the corresponding quantum channel $\Phi_{p_0}$ reduces its coherent information due to the entropy exchange: the information is hidden in the environment. This highlights the key role of the inaccessible environment in our model for implementing the quantum vault. As the system interacts with the environment, neither Alice nor Eve can obtain the information stored in its initial state.

Conveniently enough, further evolution of the system in the non-Markovian regime allows for partial o complete information recovery. In the optimal case [Fig \ref{fg:caps} (a)] the information is fully restored, so the coherent information reaches its maximal value at the output:
\begin{equation}\label{eq:Icmax}
 I_c(\rho_A')=I_c(\rho_A) =s(\rho_A)\,.
\end{equation}

This way, it is shown that in the optimal scenario, $\Lambda_t$ allows for a quantum private channel in the limit $p_0\longrightarrow 0$, after performing quantum data hiding during the evolution, maximally at $p_0=p_{\rm min}$. The previous results can be expressed by means of another information quantity obtained from a different partition of the extended system $ARE$, namely, the \emph{loss} given by
\begin{equation}
\begin{split}
  L(\rho,\Phi)&=S(\rho_R) + S(\rho_E') - S(\rho_{RE}')\\
  &= S(\rho_A) + S(\rho,\Phi) -S(\rho_A')\,.
\end{split}
\end{equation}

Notice that it corresponds to the quantum mutual information between the input and the environment. The loss is non-negative, and it is null only if the state $\rho_{RE}'$ is separable, i.e. in the absence of correlations between the $R$ and $E$, meaning that no information has leaked into the environment. Indeed, it has been shown that $L(\rho,\Phi)=0$ is equivalent to Eq. \eqref{eq:Icmax}, allowing for a completely private channel \cite{Holevo,loss}. This is attained in our optimal case, where all the information has flown back from the environment to the system at the end of the evolution.

\subsection*{QV protocol in different scenarios}
\begin{figure*}[!t]
\centering 
\includegraphics[width=0.65\textwidth]{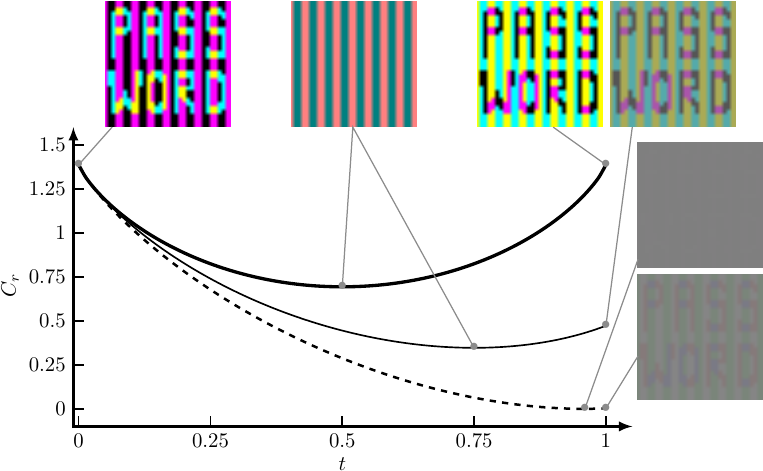}
\caption{Implementation of the quantum vault protocol to store an image, using the noisy map $\Lambda_t$ in three different scenarios. Each qudit in the 4-dimensional canonical basis is associated with a specific color of the cyan-magenta-yellow-black (CMYK) model. At the input stage, Alice stores a message as an image whose pixels have one of the base colors. As the system evolves, distinguishability between different pairs of states in the basis is reduced, which is reflected by a decrease in the quantum mutual information (which quantifies the capacity of a quantum channel to convey a classical message). At $t=1-p_{\rm min}$, the capacity is minimal, rendering the message of the image unreadable. Non-Markovian dynamics in the final stage of the evolution lead to an increase in the quantum mutual information with the respective revival of distinguishability, allowing Alice to retrieve her information. In the simplified scenario where $p_1=p_2=0$ (thick line), a full recovery is possible, even if the final state of each pixel is different from their initial state (colors are swapped in the retrieved image). In the uniform scenario $p_1=p_2=p_3=(1-p_0)/3$ (thin line), there is a partial revival of distinguishability, allowing one to read the message. Finally, the unrestricted scenario with $s=4$ (dashed line), where permutations between all the cores can occur, enables complete encryption of the image at the minimum, at the expense of a low recovery in the output.}\label{fg:qv_sp}
\end{figure*}

The implementation of the QV in Sec. \ref{sec:qvdes} considers the optimal scenario in the $s=2$ case, where phase permutations can only occur simultaneously on both subsets of cores ($p_1=p_2=0$ and $p_3=1-po$). With this restriction, the relative entropy of coherence increases back to its initial value, allowing a complete information backflow at the end of the evolution. In Fig. \ref{fg:qv_sp}, we compare how the QV performs in this optimal case and more unrestricted scenarios. The information backflow in the former case is reflected by the output image, all of whose pixels are in a pure color of the CMYK base, just like the image initially stored by Alice. In the uniform case $p_{i\neq 0}=(1-p_0)/3$ (thin filled curve) evolution until $p_0=1/4$ put the pixels in the same two mixtures as before. However, their state is not pure at the end of the evolution so Alice can recover the information only partially. Finally, the more realistic case $s=4$, where permutations can occur between all the four cores (dashed curve), enables full encryption of Alice's message, with all the base states evolving to the same mixed state, or equivalently, all the pixels in the image becoming gray. This comes at the cost of a very poor recovery at the end of the evolution, reflected by a barely readable message in the output image. This is related to the fact that the value of $p_\text{min}$ in \eqref{eq:pmin} is inversely proportional to the number of available permutations affecting the evolution.

\end{document}